\documentclass{webofc}
\usepackage[varg]{txfonts}   % Web of Conferences font
\usepackage{comment}

\begin{document}

\title{Pressure profiles of distant Galaxy clusters with {\em Planck}-SPT data}

\author{\firstname{Filippo} \lastname{Oppizzi}\inst{1}\fnsep\thanks{\email{filippo.oppizzi@gmail.com}} \and 
	\firstname{Federico} \lastname{De Luca}\inst{1} 
	\and
	\firstname{Hervé} \lastname{Bourdin}\inst{1}
	\and
	\firstname{Pasquale} \lastname{Mazzotta}\inst{1}
	\and
	\ \lastname{the CHEX-MATE collaboration}
        % etc.
}

\institute{Universit\`a degli studi di Roma `Tor Vergata',  Via della ricerca scientifica, 1; 00133 Roma, Italy          }

\abstract{%
  We present a full set of numerical tools to extract Galaxy Cluster pressure profiles from the joint analysis of {\em Planck} and South Pole Telescope (SPT) observations.
   Pressure profiles are powerful tracers of the thermodynamic properties and the internal structure of the clusters. Tracing the pressure over the cosmic times allows to constraints the evolution of the cluster structure and the contribution of astrophysical phenomena.
   SPT and Planck are complementary to constrain the cluster structure at various spatial scales. The SPT cluster catalogue counts 677 cluster candidates up to redshift 1.7, it is a nearly mass limited sample, an ideal benchmark to test cluster evolution.
   We developed a pipeline to first separate the cluster signal from the background and foreground  components and then jointly fit a parametric profile model on a combination of Planck and SPT data. We validate our algorithm on a sub-sample of six clusters, common to the SPT and the CHEX-MATE catalogues, comparing the results with the profiles obtained from X-ray observations with XMM-\textit{Newton}. 
}
\maketitle
\section{Introduction}
\label{intro}
Galaxy clusters form from the largest gravitational overdensities in the density field and track the evolution of the Universe at the largest scales \citep{2012ARA&A..50..353K}.
A large amount of ionised hot gas lies in hydrostatic equilibrium in the potential well of the Dark Matter (DM) halos of the clusters.
As a consequence of the tight correlation between the gas and the DM density field, the scale invariance of the latter reflects in the self-similarity of the pressure profiles.
The intra-cluster gas pressure is thus a sensitive tracer of the internal structure of galaxy clusters.\\
The Sunyaev-Zeldovic effect (SZ) arises from the interaction of the free electrons in the ionised gas with the Cosmic Microwave Background (CMB) photons.
It is a direct tracer of the gas pressure and is independent of the redshift. \citep{1972CoASP...4..173S}.
The gas also emits bremsstrahlung radiation in X-ray, providing a complementary probe of the cluster structure that is sensitive to different density regimes.\\
In this work, we present a pipeline to extract cluster pressure profiles from a combination of {\em Planck} and the South Pole Telescope (SPT) data.
The high sensitivity of Planck and the SPT high resolution makes the two instruments complementary in constraining the cluster structure from the core to the outskirts.
We compare the results from this SZ analysis with the profiles derived from X-ray measurements.
We analyse six clusters extracted from the sample of the Cluster HEritage project with XMM-Newton-Mass Assembly and Thermodynamics at the Endpoint of structure formation (CHEX-MATE) \citep{CHEX-MATE}.\\
In section \ref{sec:data} we present the data-sets and the cluster sample, in section \ref{sec:meth} we describe the analysis techniques and in section \ref{sec:res} we present our results. Finally, section \ref{sec:concl} is dedicated to the conclusions.

\section{Data-sets and Cluster Sample}
\begin{comment}
\begin{figure*}
    \centering
    \includegraphics[width=0.48\textwidth]{zMproc.png}
        \includegraphics[width=0.48\textwidth]{zdproc.png}
    \caption{Right: SPT cluster sample in the $z-D_{500}$ plane. Left: CHEX-MATE cluster sample in the $z-M_{500}$ plane.  In both panels the blue circles mark the six cluster analysed in this work that represent the overlap between the two catalogues. }
    \label{fig:sample}
\end{figure*}
\end{comment}
\label{sec:data}
SPT and {\em Planck} represent the state of the art of cosmological surveys at millimetre and sub-millimetre wavelengths. 
Combining the information from both data-sets it is possible to solve the inner core of all clusters in the SPT catalogue and at the same time to map the faint peripheries with high precision.\\
We use the public data from the {\em Planck} High-Frequency Instrument (HFI) \citep{2020A&A...641A...3P}. The six sky maps with nominal frequencies $100$, $143$, $217$, $353$, $545$ and $857 GHz$ and resolution 9.66, 7.22, 4.90, 4.92, 4.67, 4.22 arcmin respectively.\\
SPT observes the microwave sky in three frequency bands centred at 95, 150 and 220 GHz with 1.7, 1.2, and 1.0 arcmin resolution, respectively over an area of 2540 $\deg^2$ \citep{2011PASP..123..568C}.
Thanks to its resolution it can solve the core of distant galaxy clusters, inaccessible by {\it Planck} \citep{2015ApJS..216...27B}.
In this work, we use the public SPT maps\footnote{\url{https://pole.uchicago.edu/public/data/chown18/index.html}} that are convolved with a common Gaussian beam with 1.75 arcmins FWHM.\\
CHEX-MATE is a program of X-ray observations with the XMM-\textit{Newton} satellite of a selection of 118 galaxy clusters of the {\em Planck} PSZ2 catalogue designed to be unbiased and signal-to-noise-limited \citep{CHEX-MATE}.
To test our technique we use a set of clusters representing the overlap between the SPT and the CHEX-MATE catalogue. We use the X-ray counterparts as a benchmark for the pressure profile derived from our analysis of millimetric data. 
%Fig. \ref{fig:xb} shows the Xray surface brightness maps of the six clusters.  
The sample covers a redshift range from $z=1.6$ to $z=0.6$, with angular dimension from $r_{500}=2.8$ to $r_{500}=7.9$ arcmin.\\
\section{Method}
\label{sec:meth}
\begin{figure*}
\centering
\includegraphics[width=\textwidth,clip]{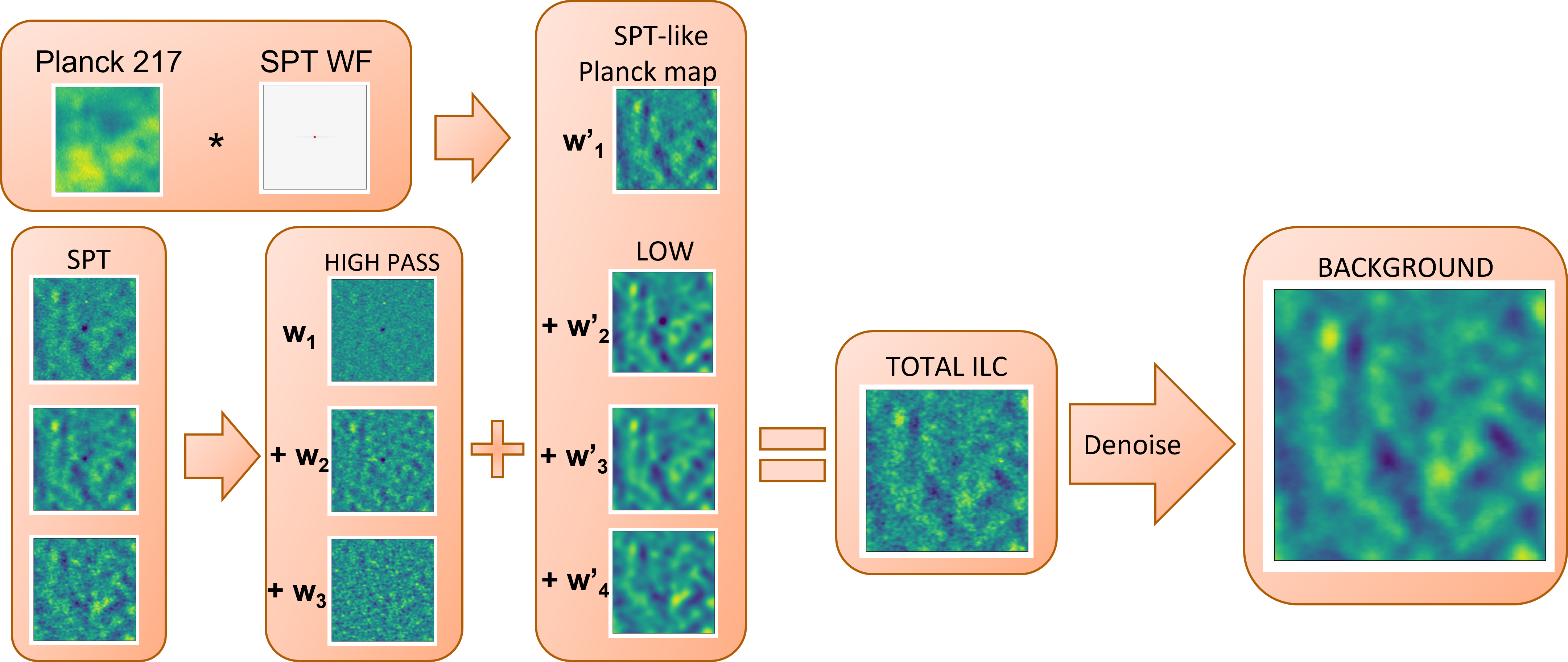}
%\vspace*{5cm}       % Give the correct figure height in cm
\caption{Flow chart that describe the SPT component separation pipeline described in section \ref{sec:sptback}}
\label{fig:flowsptback}       % Give a unique label
\end{figure*}
\label{sec:planckback}
We first perform a component separation to remove the contaminant emissions on the data from the two instruments separately and then we fit the pressure profiles on both data-sets.
We refer to  \cite{inprep} for detailed description of the method.\\
\subsection{Processing of {\it Planck} data}
We project the full sky {\em Planck} maps on tiles of $512\times512$ pixels, with the resolution of 1 arcmin/pixel centred on the cluster.
Each image is then processed with the technique developed in \cite{2017ApJ...843...72B} to isolate the cluster signal from the background and foreground components.
The technique consists of a four components fit taking into account the cluster SZ signal, the primary CMB anisotropies, the Galactic Thermal Dust and a correction for the intra-cluster Dust emission.
We use a wavelet thresholding algorithm to obtain the spatial templates of the CMB and the Galactic dust directly from the 857 GHz and the 217 GHz {\it Planck} channels respectively. 
To model the thermal dust spectral energy distribution we use the two-temperature grey body proposed in \cite{2015ApJ...798...88M}.
We compute the correction for intra-cluster dust assuming a  Navarro-Frenk-White profile (NFW) and a grey body spectral energy distribution (SED).
All the templates are then convolved with the proper beams to match the resolution of each channel, fitted and subtracted to the raw data to obtain the cleaned maps where we will perform the cluster fit.
\subsection{Processiong of SPT data}
\label{sec:sptback}
As for {\it Planck} we  select patches of $512\times512$ pixels around the XMM cluster centre, but with resolution of 0.5 arcmin/pixel instead of 1 arcmin/pixel.
We do not use the same technique applied in the {\it Planck} data reduction, since SPT lacks the high-frequency channels necessary to model the thermal dust.
On the other hand, SPT is insensitive to the large spatial scales where the dust emission is relevant, our clusters are far from the dusty galactic plane,  and SPT observe at low frequencies where the dust SED is low. Thus, we do not expect to observe significant dust contamination, and we do not find any evidences of it in our analysis. Therefore, we focus on the modelling of the CMB signal using a linear combination (LC) of the three SPT channels and the 217GHz channel of {\em Planck} that we re-project on a map with the same characteristic of the SPT ones.
We compute the weights of the LC minimising the variance with respect to a signal with a flat SED ({\em i.e.} the CMB) and simultaneously to null the non-relativistic SZ component.
We split then each map in a low-pass filtered map matching the {\em Planck} 217 resolution, and in a high pass filtered map containing the residual small scale features. 
We then perform the LC separately on the low-passed maps and high-passed maps, including the {\em Planck map} in the combination of the large scales.
We obtain two maps representing the large and small scales CMB fluctuations, that we re-combine into a single map.
Finally, we denoise the CMB template with a wavelet thresholding algorithm.
Fig. \ref{fig:flowsptback} shows the flow chart of the process just described.
\begin{figure*}
    \centering
    \includegraphics[width=\textwidth]{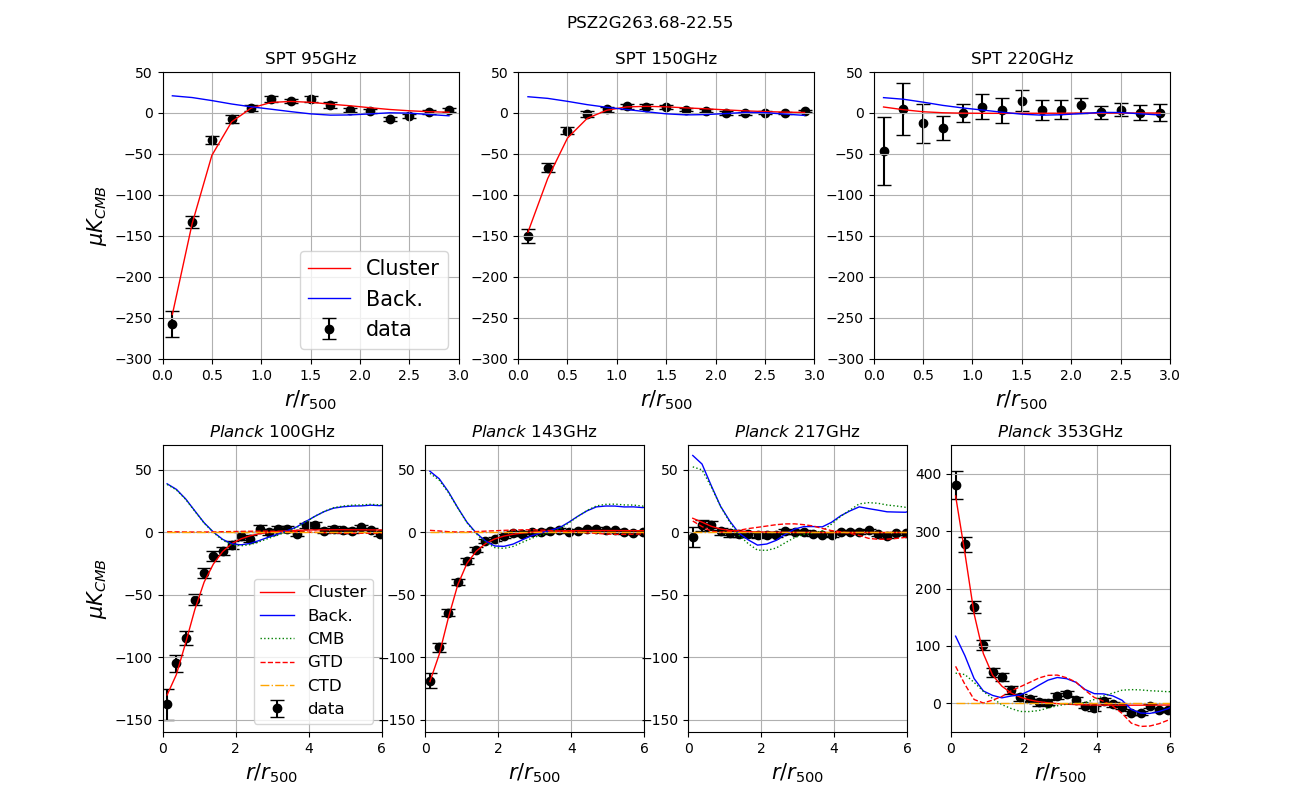}
    \caption{Intensity profiles around PSZ2G259.98-63.43 for the SPT channels ( upper panels) and {\em Planck} channels (lower panels). Black dots represent the radial average on the cleaned maps, the red line is the cluster profile, the blue line is the background. For {\em Planck} the dotted cyan line is the CMB, the dashed red line is the GTD and the dash-dotted orange line is the cluster dust.   }
    \label{fig:freqprof263}
\end{figure*}
\subsection{Joint fit}
\label{sec:fit}
We combine the SPT and {\em Planck} data in a joint fit of a cluster thermal SZ template derived from the projection along the line of sight of a generalized NFW (gNFW) pressure profile:
\begin{equation}
    P(r)=P_{0}\times \frac{P_{500}}{x^\gamma(1+x^\alpha)^{(\beta-\gamma)/\alpha}},
    \label{eq:pressure}
\end{equation}
where $x=c_{500}r/r_{500}$, $r_{500}$ is the radius where the density is 500 times the critical density of the Universe at the cluster redshift, $c_{500}$ is the concentration parameter and  $\alpha,\beta$ and $\gamma$ are the slopes at $r\sim r_{500}/c_{500}$,  $r\gg r_{500}/c_{500}$ and $r\ll r_{500}/c_{500}$ respectively.
The template is then converted in SZ emissivity and convolved with the beam and the SPT window function to build a multi channel model of the cluster signal.
We compute the likelihood on the radially averaged profiles of the cleaned data and the model. 
We use circular bins of width $0.2r_{500}$ for SPT, and of 2 arcmin for {\em Planck}, up to $3 r_{500}$ for SPT and $6r_{500}$ for {\em Planck}.
We assume a Gaussian Likelihood and we perform a MCMC fit with the Cobaya MCMC sampler \citep{2021JCAP...05..057T}. 
For {\em Planck} channels, we obtain the covariance matrices from Monte-Carlo simulations of the noise in the surroundings of the cluster.
For SPT channels instead, we use the sample covariance of the noise maps provided by the SPT collaboration \citep{2018ApJS..239...10C}. 
We try different combinations of parameters to optimise the convergence of our chain.
We conclude to leave free to vary the amplitude $P_{0}$ and the inner and outer slopes $\gamma$ and $\beta$ and to keep the concentration parameter and the intermediate slope fixed to their universal values identified by \cite{2010A&A...517A..92A} $c_{500}=1.177$ and $\alpha=1.0501$.
\subsection{X-ray analysis}
\label{sec:Xana}
We based our X-ray analysis on the observations of the European Photon Imaging Camera (EPIC) public data of the XMM-\textit{Newton} telescope. 
We assume a background model taking into account the quiescent particle background, the cosmic X-ray background, and the thermal emission associated with our Galaxy. We threat flares and high-background periods as in \cite{Bourdin2013}.
We estimate the X-ray thermodynamical profiles as done in \cite{2017ApJ...843...72B}. 
We use the expressions introduced in \cite{Vikhlinin2006} to parametrise the emission measure $[n_p n_e](r)$ and temperature $T(r)$.
We then integrate the templates along the line of sight and we fit them  to the background-subtracted X-ray observable profiles.
We refer to \cite{Delucainprep} for a complete description of our X-ray data reduction technique.

\section{Results}
\begin{figure*}
    \centering
    \includegraphics[width=0.48\textwidth]{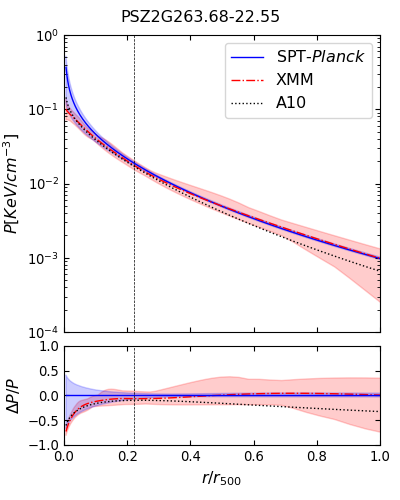}
        \includegraphics[width=0.48\textwidth]{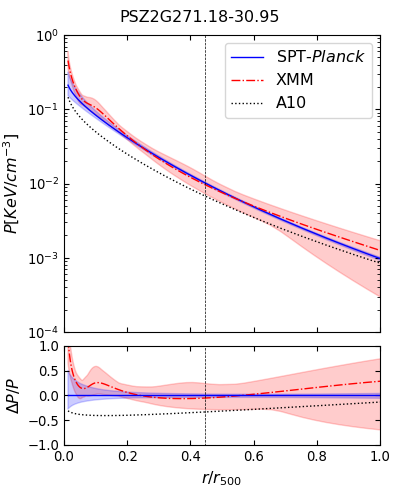}
    \caption{Profiles from XMM-\textit{Newton} data (dash dotted red lines) and  from SPT-Planck joint fit (blue lines). The shaded regions correspond to the $68\%$ credible intervals. The dotted black lines show the universal profile from \cite{2010A&A...517A..92A}. The dashed lines indicate the FWHM of the SPT beam.}
    \label{fig:presprof}
\end{figure*}
\label{sec:res}
We show in Fig. \ref{fig:freqprof263} the intensity profile of the estimated cluster signal and the contaminant components at each frequency for one of the clusters in our sample, compared with the cleaned data. 
We compare the pressure profile derived from the SZ maps with the expectations from X-ray observations.
Since X-ray and SZ data probe the cluster structure at different regimes, we limit our comparison to the overlapping region between the two, {\em i.e.} inside $r_{500}$.
We show a preliminary comparison between the two fits on two clusters of our sample in Fig. \ref{fig:presprof}.
The agreement is remarkable: the credibility intervals overlap along the whole radius range tested. 
Some degree of divergence can be observed in the very inner regions that are not solved by SPT.
The agreement between the two analysis confirms the consistency of our techniques.

%-------------------------------------
\section{Conclusions}
\label{sec:concl}
We presented the preliminary results of the measurement of the pressure profile of galaxy clusters from the joint analysis of SPT and {\em Planck} data.
We developed two independent data reduction pipelines to fully exploit the potential of each instrument.
We fit a spherically symmetric gNFW model of the profiles, processed by the instrumental response of each channel, to efficiently exploit the information content of each band.
We applied our algorithm to a sample of six galaxy clusters representing the intersection between the SPT and CHEX-MATE catalogues. 
Here we presented the preliminary results on two clusters and we refer to \cite{inprep} for the complete analysis.
We compared the profiles derived from the SPT-{\em Planck} joint fit to the ones obtained from the XMM-\textit{Newton} data.
We found a good agreement between the results of the two measurements. 
With the confirmation of the consistency of our method in hand, we will expand our analysis to a larger number of SPT clusters to investigates the structure and evolution of the profiles through cosmic time.
\bibliography{SPTxPlanck}

\end{document}